\documentclass[twocolumn]{pasj01}

\RequirePackage{mathpazo}
\RequirePackage[T1]{fontenc}

\usepackage{comment}
\usepackage{lineno}

\def\commenta{$^*$}

\newcounter{author}
\def\authorcount#1#2{\refstepcounter{author}\label{#1}
                     \altaffiltext{\ref{#1}}{#2}}

\Received{$\langle$reception date$\rangle$}
\Accepted{$\langle$acception date$\rangle$}
\Published{$\langle$publication date$\rangle$}

\begin{document}
\SetRunningHead{Y. Tampo et al.}{PNV J00444033+4113068 2021 superoutburst}

\title{PNV J00444033+4113068: early superhumps with 0.7 mag amplitude and non-red color}

\author{
    Yusuke~\textsc{Tampo}\altaffilmark{\ref{affil:Kyoto}*},
    Keisuke~\textsc{Isogai}\altaffilmark{\ref{affil:KyotoOkayama}}$^,$\altaffilmark{\ref{affil:Kyoto}}$^,$\altaffilmark{\ref{affil:MuSCAT}},
    Naoto~\textsc{Kojiguchi}\altaffilmark{\ref{affil:Kyoto}},
    Makoto~\textsc{Uemura}\altaffilmark{\ref{affil:HAO}},
    Taichi~\textsc{Kato}\altaffilmark{\ref{affil:Kyoto}},
    Tam\'as~\textsc{Tordai}\altaffilmark{\ref{affil:Trt}},
    Tonny~\textsc{Vanmunster}\altaffilmark{\ref{affil:Van1}}$^,$\altaffilmark{\ref{affil:Van2}},
    Hiroshi~\textsc{Itoh}\altaffilmark{\ref{affil:Ioh}},
    Pavol~A.~\textsc{Dubovsky}\altaffilmark{\ref{affil:Vih}}, 
    Tom\'a\v{s}~\textsc{Medulka}\altaffilmark{\ref{affil:Vih}}, %
    Yasuo~\textsc{Sano}\altaffilmark{\ref{affil:san}}$^,$\altaffilmark{\ref{affil:san2}}$^,$\altaffilmark{\ref{affil:san3}},
    Franz-Josef~\textsc{Hambsch}\altaffilmark{\ref{affil:ham1}}$^,$\altaffilmark{\ref{affil:ham2}}$^,$\altaffilmark{\ref{affil:dfs}}, 
    Kenta~\textsc{Taguchi}\altaffilmark{\ref{affil:Kyoto}},
    Hiroyuki~\textsc{Maehara}\altaffilmark{\ref{affil:NAOJOkayama}}$^,$\altaffilmark{\ref{affil:KyotoOkayama}}, 
    Junpei~\textsc{Ito}\altaffilmark{\ref{affil:Kyoto}}, and    
    Daisaku~\textsc{Nogami}\altaffilmark{\ref{affil:Kyoto}}
}

\authorcount{affil:Kyoto}{
     Department of Astronomy, Kyoto University, Kyoto 606-8502, Japan}
\email{$^*$tampo@kusastro.kyoto-u.ac.jp}

\authorcount{affil:KyotoOkayama}{
     Okayama Observatory, Kyoto University, 3037-5 Honjo, Kamogatacho,
     Asakuchi, Okayama 719-0232, Japan}

\authorcount{affil:MuSCAT}{
    Department of Multi-Disciplinary Sciences, Graduate School of Arts and Sciences, 
    The University of Tokyo, 3-8-1 Komaba, Meguro, Tokyo 153-8902, Japan}

\authorcount{affil:HAO}{
Hiroshima Astrophysical Science Center, Hiroshima University, Kagamiyama 1-3-1, Higashi-Hiroshima 739-8526, Japan}

\authorcount{affil:Trt}{
     Polaris Observatory, Hungarian Astronomical Association,
     Laborc utca 2/c, 1037 Budapest, Hungary}

\authorcount{affil:Van1}{
    Center for Backyard Astrophycis Belgium, 
    Walhostraat 1a, B-3401 Landen, Belgium}

\authorcount{affil:Van2}{
    Center for Backyard Astrophycis Extremadura, 
    e-EyE Astronomical Complex, 
    ES-06340 Fregenal de la Sierra, Spain}
    
\authorcount{affil:Ioh}{
     Variable Star Observers League in Japan (VSOLJ),
     1001-105 Nishiterakata, Hachioji, Tokyo 192-0153, Japan}
     
\authorcount{affil:Vih}{
     Vihorlat Observatory, Mierova 4, 06601 Humenne, Slovakia}

\authorcount{affil:san}{
    Variable Star Observers League in Japan (VSOLJ), Nishi juni-jou minami 3-1-5, Nayoro, Hokkaido, Japan}

\authorcount{affil:san2}{
    Observation and Data Center for Cosmosciences, Faculty of Science, Hokkaido University, Kita-ku, Sapporo, Hokkaido 060-0810, Japan}

\authorcount{affil:san3}{
Nayoro Observatory, 157-1 Nisshin, Nayoro, Hokkaido 096-0066, Japan}

\authorcount{affil:ham1}{
    Groupe Européen d’Observations Stellaires (GEOS), 23 Parc de Levesville, 28300 Bailleau l’Evêque, France}
\authorcount{affil:ham2}{
    Bundesdeutsche Arbeitsgemeinschaft für Veränderliche Sterne (BAV), Munsterdamm 90, 12169 Berlin, Germany}

\authorcount{affil:dfs}{
    Vereniging Voor Sterrenkunde (VVS), Oostmeers 122 C, 8000 Brugge, Belgium}
    
\authorcount{affil:NAOJOkayama}{
     Subaru Telescope Okayama Branch Office, National Astronomical Observatory of Japan, 
     National Institutes of Natural Sciences, 3037-5 Honjo, Kamogata, Asakuchi, Okayama 719-0232, Japan}



\maketitle

\begin{abstract}

In the first days of WZ Sge-type dwarf nova (DN) outbursts, the 2:1 resonance induces a spiral arm structure in the accretion disk, which is observed as early superhumps in optical light curves.
This paper reports our optical observations of an eclipsing WZ Sge-type DN PNV J00444033+4113068 during its 2021 superoutburst with the 3.8m Seimei telescope and through VSNET collaboration.
The eclipse analysis gave its orbital period as 0.055425534(1) d.
Our observations confirmed early superhumps with an amplitude of 0.7 mag, the largest amplitude among known WZ Sge-type DNe.
More interestingly, its early superhumps became the reddest around their secondary minimum, whereas other WZ Sge-type DNe show the reddest color around the early superhump maximum. 
The spectrum around the peak of the outburst showed the double-peaked emission lines of He II 4686\AA~ and H$\alpha$ with a peak separation of $\ge 700$ km/s, supporting a very high-inclination system.
With the early superhump mapping, the unique profile and color of the early superhump of PNV J00444033+4113068 are successfully reproduced by the accretion disk with vertically extended double arm structure.
Therefore, the large amplitude and unique color behavior of the early superhumps in PNV J00444033+4113068 can be explained by the 2:1 resonance model along with other WZ Sge-type DNe.

\end{abstract}


\section{Introduction}
\label{sec:1}

Dwarf novae (DNe) are accreting white dwarf (WD) binaries that possess an accretion disk and show recurrent outbursts (see \cite{war95book, hel01book}).  
It is widely accepted that the mechanism of DN outbursts is explained by the tidal-thermal disk instability model [for a review, see \citet{osa96review}].
WZ Sge-type DNe form a subclass in DNe.
Their outbursts are characterized by early superhumps observed for the first 5-10 days of the outburst, which is a double-wave variation with a period almost identical to the orbital period of the system \citep{ish02wzsgeletter,kat15wzsge}.
While most of WZ Sge-type DNe show early superhumps with the amplitude less than 0.05 mag \citep{kat15wzsge,kat22WZSgecandle},  high-inclination systems show the amplitude larger than 0.1 mag [e.g., OV Boo \citep{pat08j1507}, V455 And \citep{mat09v455and, Pdot}, ASASSN-18do (vsnet-alert 21921\footnote{http://ooruri.kusastro.kyoto-u.ac.jp/mailarchive/vsnet-alert/21921})].
In addition, the multi-color observations of early superhumps showed the reddest color around the early superhump maxima \citep{mat09v455and, nak13j0120, iso15ezlyn, ima18HVVirJ0120}.  
These results have suggested that the early superhumps can be explained by the orbital rotational effect of the outer disk with a non-axisymmetric vertical structure \citep{nog97alcom, mat09v455and, ima18HVVirJ0120}.

The theoretical understanding of early superhumps is considered to be the occurrence of the 2:1 resonance between the secondary star and the Keplerian accretion disk, resulting in vertical deformation of the accretion disk and the appearance of a double spiral arm pattern \citep{lin79lowqdisk, osa02wzsgehump, kun04SHSPH, kun05earySHSPH}.
On the other hand, ordinary superhumps are induced by the eccentric disk through the 3:1 resonance \citep{whi88tidal,osa89suuma, hir90SHexcess}.
As the 2:1 resonance suppresses the growth of the 3:1 resonance \citep{lub91SHa}, early superhumps are always observed before ordinary superhumps.

Observational evidence of the spiral arm structure in WZ Sge-type DNe was first found in the WZ Sge 2002 superoutburst \citep{bab02wzsgeletter,kuu02wzsge}.
Applying Doppler tomography with He II 4686\AA, a double-arm structure in the accretion disk was deduced.
Another observational approach of studying a disk structure is to model the profile of early superhumps \citep{mae07bcuma, uem12ESHrecon}.
By modeling the multi-color early superhump profiles with self-occultation of the vertically extended disk, called "early  superhump mapping", \citet{uem12ESHrecon,nak13j0120} revealed the double arm spiral structure in the accretion disk, highlighting the occurrence of the 2:1 resonance in the accretion disk of WZ Sge-type DNe.
However, due to the limited samples, the diversity of the disk structure is still not investigated.

PNV J00444033+4113068\footnote{http://tamkin1.eps.harvard.edu/unconf/followups/J00444033+4113068.html} (= AT 2021aaxp, hereafter PNV J0044) was discovered as an M31 classical nova candidate
by Koichi Itagaki at 16.5 mag on 2021-10-09.4579.
However, double-peaked emission lines of H$\alpha$ and He II 4686\AA~ were detected in the follow-up spectroscopic observation, classifying  PNV J0044  as a foreground large-amplitude DN rather than an M31 classical nova \citep{tag21PNVJ0044}.
Later photometric observations detected  early superhumps with the amplitude of 0.7 mag, confirming PNV J0044 as a WZ Sge-type DN (vsnet-alert 26319 \footnote{http://ooruri.kusastro.kyoto-u.ac.jp/mailarchive/vsnet-alert/26319}).
The quiescence counterpart is likely V$=22.278$ mag according to the Revised LGGS UBVRI photometry of the M31 and M33 stars catalog \citep{mas16M31M33stars}.

In this paper, we present our optical observations and analyses of PNV J0044 during its 2021 outburst.
Section \ref{sec:2} presents the overview of our observations of the superoutburst,  and Section \ref{sec:3} shows the results of our analysis. 
We discuss the properties  of PNV J0044 in Section \ref{sec:4} and give the summary of this paper in Section \ref{sec:5}.

\section{Observations and Analysis}
\label{sec:2}

\subsection{photometric observations}
Our time-resolved CCD photometric observations of PNV J0044 were carried out by the Variable Star Network (VSNET) collaborations \citep{VSNET}. 
We also performed the simultaneous $g$-, $r$- and $i$-band  photometry with the TriColor CMOS Camera and Spectrograph (TriCCS\footnote{http://www.kusastro.kyoto-u.ac.jp/$	\textasciitilde$kazuya/p-triccs/index.html}) mounted on the 3.8m Seimei telescope \citep{kur20seimei} at Okayama Observatory of Kyoto University.
The instruments for our photometric observations and our observation logs are summarized in tables E1 and E2 \footnote{Tables E1 and table E2 are available only on the online edition as Supporting Information. }, respectively. 
All the observation epochs in this paper are described in the Barycentric Julian Day (BJD). 
VSNET observations were unfiltered, and the zero point of these data was adjusted to the observations by T. Vanmunster for our period analysis.
For the magnitude calibration of TriCCS data, the AAVSO comparison star 000-BNN-553 (= Gaia DR3 369265315130536960) with $g = 16.039(3)$, $r = 15.621(3)$, and $i = 14.473(1)$ at $(\alpha, \delta)_{\rm J2000.0} =$ (\timeform{00h44m44s.96}, $+$\timeform{41D13'12''.9}) (Pan STARRS DR1; \cite{panstarrs1}) was adopted.  
We also extracted photometric survey data from the Zwicky Transient Facility (ZTF; \cite{ZTF}) alert broker Lasair \citep{lasair}  and the Asteroid Terrestrial-impact Last Alert System (ATLAS; \cite{ATLAS}) to examine the global light curve profiles.
These survey data were not included in our period analysis.

The phase dispersion minimization (PDM;  \cite{PDM}) method was applied for period analysis of the superhumps in this paper. 
The 1$\sigma$ errors for the PDM analysis was determined following \citet{fer89error, pdot2}.
Before period analysis, the global trend of the light curve was removed by subtracting a smoothed light curve obtained by locally weighted polynomial regression (LOWESS: \cite{LOWESS}).

\subsection{spectroscopic observations}

We performed our spectroscopic observation of PNV J0044 on BJD 2459497.21 using the fiber-fed integral field spectrograph (KOOLS-IFU; \cite{mat19koolsifu}) mounted on Seimei telescope \citep{kur20seimei}.
We applied  VPH-blue as a grism, which has a resolution of $R\sim 500$ and a wavelength coverage of 4,200-8,000\AA. 
Our data reduction was performed using IRAF\footnote{IRAF is distributed by the National Optical Astronomy Observatories, which are operated by the Associations of Univesities for Research in Astronomy, Inc., under cooperative agreement with the National Science Foundation.} in the standard manner (bias subtraction, flat fielding, aperture determination, spectral extraction, wavelength
calibration with arc lamps, and flux calibration with a standard star).
The preliminary result was already reported by \citet{tag21PNVJ0044}.

\section{Results}
\label{sec:3}

\subsection{Overall light curve during the superoutburst}
\label{sec:3.1}

\begin{figure}[tbp]
 \begin{center}
  \includegraphics[width=80mm]{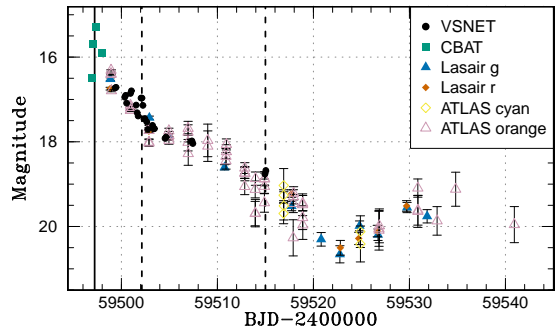}
 \end{center}
 \caption{
 The light curve of PNV J0044 during the 2021 superoutburst.
 Black filled circles, green filled squares, blue filled triangles, orange filled diamonds, pink open triangles, and open yellow diamonds represent data from VSNET, CBAT, ZTF Lasair $g$, ZTF Lasair $r$, ATLAS orange, and ATLAS cyan band, respectively.
 The VSNET data are binned in 0.1 d.
 The vertical solid and dashed lines show the epoch of our spectroscopic observation with KOOLS-IFU and photometric observations with TriCCS, respectively.}
 \label{fig:2021lc}
\end{figure}

Figure \ref{fig:2021lc} shows the global light curve of PNV J0044 during the superoutburst in 2021.
Its peak magnitude is $\sim15.3$ mag on BJD 2459497, and hence the outburst amplitude reached $\sim7$ mag.
Our time-resolved observations between BJD 2459500 and 2459504 showed clear early superhumps, whereas on BJD 2459507, the variation profile was featureless due to the low S/N of the data.
The slope of the outburst decay became gentler around BJD 2459506, which can be attributed to the alternation of early and ordinary superhumps \citep{kat15wzsge}.
The outburst probably ceased on BJD 2459520, and the later phase after BJD 2459525 can be a rebrightening.
As there are no upper limit observations in Lasair and ATLAS, the rebrightening profile appears to be type-A (plateau rebrightening; \cite{ima06j0137}).
We note that lacks of time-resolved observations of the later outburst phase prevent us from drawing any solid conclusions.

\subsection{eclipse and orbital period}
\label{sec:3.2}

\begin{figure*}[tbp]
 \begin{center}
  \includegraphics[width=160mm]{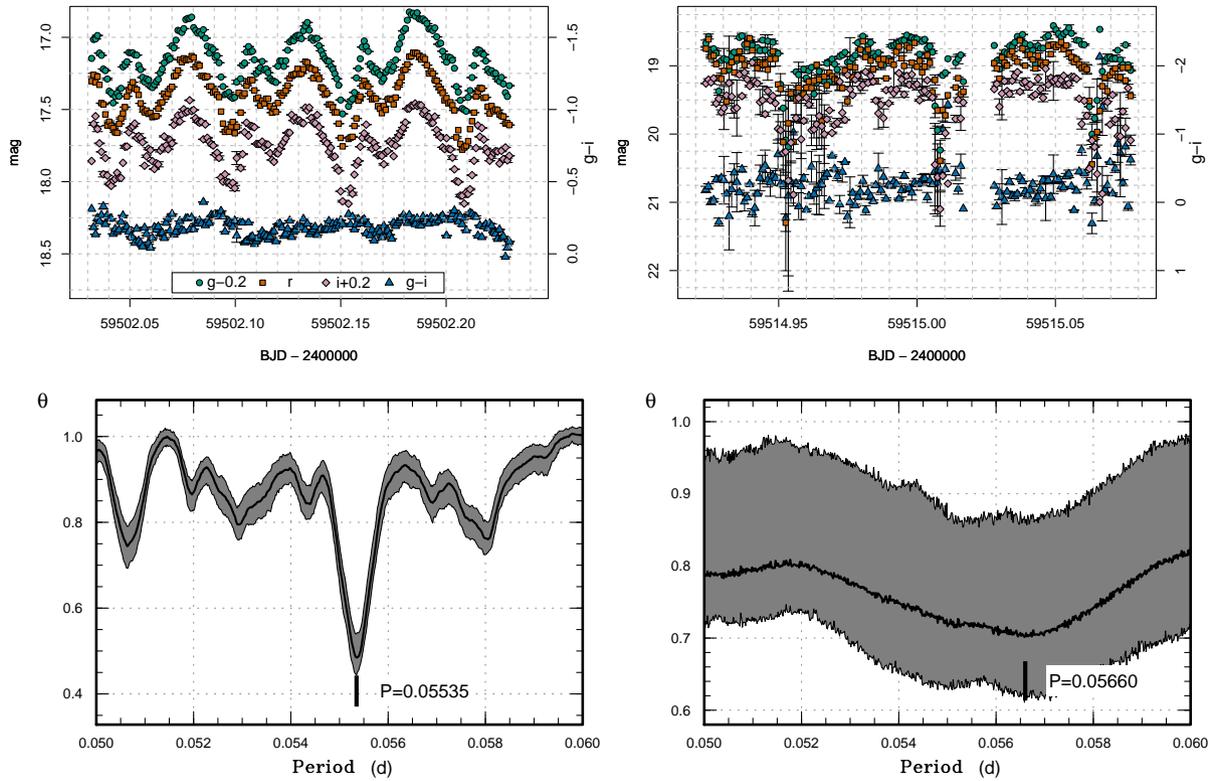}
 \end{center}
 \caption{
 Upper panels: zoomed light curve on BJD 2459502 (upper left) and 2459515 (upper right) observed with TriCCS.
 The green circles, orange squares, and pink diamonds represent the observations in $g$, $r$ and $i$ band, respectively.
 The data are binned in 0.001 d.
 The blue triangles show the $g-i$ color as well.
 Lower panels: $\theta$-diagram of the PDM analysis using data on BJD 2459500 - 2459506 (lower left) and BJD 2459514 - 2459515 (lower right). 
 The gray area represents the 90 $\%$ confidence range of $\theta$ statistics by the PDM method.}
 \label{fig:enlargeLC}
\end{figure*}

In figure \ref{fig:enlargeLC}, the enlarged light curves in the $g$, $r$ and $i$ bands on BJD 2459502 (upper left panel) and 2459515 (upper right panel) observed with TriCCS are presented.
In both panels, deep eclipses are recognized.
The depth of the eclipses was $\sim$ 0.3 mag on BJD 2459502 and $\sim1.5$ mag on BJD 2459515.
We determined the eclipse minima by fitting the Gaussian function, and then the orbital ephemeris of PNV J0044 was obtained as Equation \ref{eq:porb}.

\begin{equation}
\label{eq:porb}
    \phi_0 = \rm{BJD~} 2459500.3237(1) + 0.055425534(1) \times E
\end{equation}

The period obtained is close to the period minimum \citep{kni11CVdonor, kat22updatedSHAmethod}.
Therefore, PNV J0044 is one of the best candidates for period bouncing objects, while we cannot constrain its mass ratio due to the lack of further information such as ordinary superhump periods.

\subsection{early superhumps}
\label{sec:3.3}

In the upper left panel of figure \ref{fig:enlargeLC}, the double-peaked early superhumps are recognized on BJD 2459502.
The period of early superhumps was calculated through PDM analysis using data outside of the eclipse (lower left panel of figure \ref{fig:enlargeLC}).
The yield period is 0.05535(1) d, which is $\sim0.1\%$ shorter than the orbital period obtained from the eclipse analysis (Section \ref{sec:3.2}).
This slight difference from the orbital period is also observed in other WZ Sge-type DNe (e.g., \cite{ish02wzsgeletter}).
The superhump maxima are summarized in table E3 \footnote{Table E3 is available only on the online edition as Supporting Information.}.

The amplitude of early superhumps is $\sim 0.7$ mag including the eclipse or $\sim 0.5$ mag without the eclipse, which is the largest value for the amplitude of  early superhumps in known WZ Sge-type DNe (\cite{kat15wzsge, kat22WZSgecandle} and reference there in).
As a larger amplitude is observed in a system with higher inclination, PNV J0044 can have the largest inclination angle in WZ Sge-type DNe.

In addition, the $g-i$ color of the early superhumps became the reddest around the secondary minimum, and the peaks did not show significant redder color.
This behavior is in contradiction with other WZ Sge-type DNe such as V455 And \citep{mat09v455and} and OT J012059.6+325545 \citep{nak13j0120}, which showed the reddest color around the primary maximum of the early superhumps.
Since the redder color around the maximum of early superhumps is considered to be responsible for the vertical expansion of the outer disk \citep{mat09v455and, nak13j0120}, the early superhumps in PNV J0044 may not be explained in the same manner.

On BJD 2459515, in addition to the eclipse, PNV J0044 showed the variation with the double-peaked profile.
Our PDM analysis using data outside the eclipse yielded a period of 0.0566 (2) d for this variation, which is 1.35\% longer than the orbital period.
This variation is most likely ordinary superhumps based on the global light-curve profile; however, the superhump profile may not be prominent as this epoch corresponds to the end phase of the outburst.

\subsection{spectroscopic observation}
\label{sec:3.4}

\begin{figure}[tbp]
 \begin{center}
  \includegraphics[width=80mm]{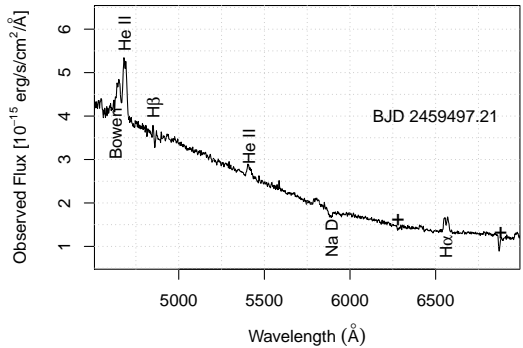}
 \end{center}
 \caption{The low-resolution spectrum of PNV J0044 obtained on BJD 2459497.21 (the orbital phase $\phi \sim $ 0.875; out of the eclipse) with KOOLS-IFU mounted on Seimei telescope.}
 \label{fig:spec}
\end{figure}

Our spectrum on BJD 2459497.21 observed with KOOLS-IFU on the Seimei telescope is presented in figure \ref{fig:spec}.
This epoch corresponds to the orbital phase $\phi \sim $ 0.875 (out of the eclipse).
The spectrum showed the blue continuum attributing to the multi-temperature disk black body and the double-peaked emission lines of H$\alpha$ and He II 4686.
He II 5411 and C $_{\rm III}$/N $_{\rm III}$ Bowen blend emission lines were detected as well.
H$\beta$ line was likely in  emission with a deep absorption core.
Na D absorption line was detected.
The peak separation of H$\alpha$ and He II 4686\AA~ is $\ge$700 km/s.
This separation is comparable to WZ Sge at its outburst peak \citep{nog04wzsgespec}, although noting that our spectrum was obtained with a low-resolution grism (R$\sim$500).
Such a large peak separation around the optical peak again supports that PNV J0044 is a high-inclination system.

The combined equivalent width (EW) of the He II 4686 and Bowen blend (-14.1\AA) was significantly larger than that of H$\alpha$ (-6.3\AA).
As in most WZ Sge-type DNe the EW of He II 4686 is weaker than H$\alpha$ (\cite{tam21seimeiCVspec} and reference therein), this result also highlights that PNV J0044 is a  high-inclination system, in which the He II emission from the heated arm structure contributes greatly to the line strength \citep{bab02wzsgeletter, mor02DNspectralatlas, tam21seimeiCVspec}.

\section{Discussion}
\label{sec:4}

As described in Section \ref{sec:3}, the early superhumps of PNV J0044 become redder around the secondary minimum rather than around the primary maximum, which is very unique color trend compared with other WZ Sge-type DNe.
Moreover, the amplitude of early superhumps is the largest among known WZ Sge-type DNe so far \citep{kat15wzsge, kat22WZSgecandle}.
Therefore, it is worth examining whether the above properties of the early superhumps of PNV J0044 can be modeled by the vertically extended spiral arm structure similar to other modeled objects (V455 And; \cite{uem12ESHrecon}, OT J012059.6+325545; \cite{nak13j0120}).
We therefore performed the early superhump mapping using the code developed in \citet{uem12ESHrecon}.
This model assumes that early superhumps are caused by vertical deformation and orbital rotation effects of the accretion disk.
The disk is assumed to radiate as blackbody and the disk temperature $T$ gradient to the disk radius $R$ to be the standard disk model ($T \propto R^{-3/4}$; \cite{sha73alphadisk}).
For this analysis, we used the simultaneous observations with TriCCS in the $g$, $r$ and $i$ bands.
The system parameters adopted for the early superhump mapping are summarized in table \ref{tab:1}.
Since the orbital period $P_{\rm orb}$ of PNV J0044 is  close to the period minimum \citep{kni11CVdonor, kat22updatedSHAmethod}, we applied 0.08 as the mass ratio $q~(=M_{\rm Secondary}/M_{\rm WD})$ assuming that PNV J0044 follows the standard evolution path of DNe.
The WD mass $M_{\rm WD}$ was adopted as 0.8 $M_\odot$, which is the typical value of the WD mass in DNe (\cite{pal22WDinCVs} and the reference therein).
For the outer disk radius $R_{\rm out}$, the 2:1 resonance radius (0.6$a$ where $a$ is the binary separation) was applied \citep{osa02wzsgehump}.
The inclination of the system $i$ was assumed to be 85$^\circ$.
This is because, as OV Boo with the early superhump amplitude of 0.27 mag is estimated to have the inclination of 83$^\circ$ \citep{pat08j1507}, the inclination of PNV J0044 is expected to be larger than OV Boo.
The innermost temperature of the accretion disk $T_{\rm in}$ was set as 150,000K, which gives the most reasonable fit.

\begin{table}
\caption{System parameters adopted in the early superhump mapping of PNV J0044.}
\centering
\label{tab:1}
\begin{tabular}{cc}
  \hline              
    Parameter   &  \\
    \hline
    \hline
    $P_{\rm orb}$ & 0.055425534 d \\
    $q$ & 0.08 \\
    $i$ & 85$^\circ$ \\
    $M_{\rm WD}$ & 0.8 $M_\odot$ \\
    $T_{\rm in}$ & 150,000 K \\
    $R_{\rm out}$ & 0.6 $a$\commenta\\
  \hline
    \multicolumn{2}{l}{\commenta $a$ is the binary separation.}\\
\end{tabular}
\end{table}

\begin{figure*}[tbp]
 \begin{center}
  \includegraphics[width=170mm]{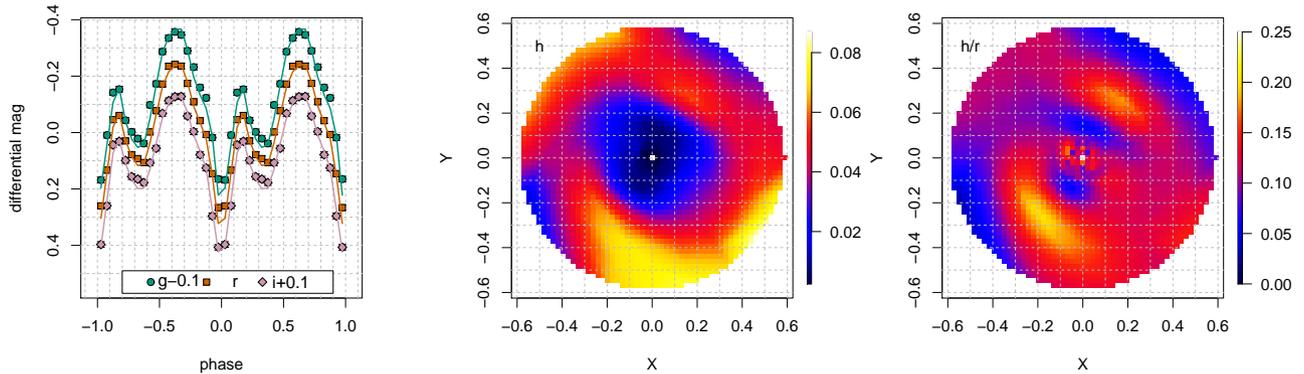}
 \end{center}
 \caption{Left panel: the phase-averaged light curves of early superhumps of PNV J0044 in the $g$ (green circles), $r$ (red squares), and $i$ (pink diamonds) bands. 
 The typical error size of 0.01 mag is also shown with the light curves.
 The differential magnitude from the mean magnitude of each band is shown.
 The solid lines represent the synthesized light curves from the early superhump mapping.
 Middle panel: the reconstructed height map of the accretion disk in height $h$  scale.
 X and Y axis, and height $h$ are normalized by the binary separation $a$.
 Right panel: the reconstructed height map of the accretion disk in the height $h$ over the radius $r$ scale.
 X and Y axis are normalized by the binary separation $a$.}
 \label{fig:eshmodel}
\end{figure*}

Figure \ref{fig:eshmodel} presents the reconstructed height map of the accretion disk of PNV J0044 (middle and right panels).
The middle panel shows the height scale normalized by the binary separation, and the right panel shows the ratio of the disk height $h$ over the radius of the disk $r$.
The synthesized light curves (solid lines in the left panel of figure \ref{fig:eshmodel}) are presented along with the phase-averaged early superhumps in the $g$ (green circles), $r$ (red squares) and $i$ (pink diamonds) bands.
As seen in the left panel, the reconstructed accretion disk well explains the observed profile and color of early superhumps of PNV J0044.
In the middle and right panels of figure \ref{fig:eshmodel}, the outermost region of the reconstructed disk shows two flaring parts in the upper left [around $(X,Y)=(-0.3,0.4)$] and lower right [around $(X,Y)=(0.1,-0.5)$] quadrants.
In addition to this, the elongated arm structure into the inner disk is recognized around $(X,Y)=(0.2,0.3)$ and $(X,Y)=(-0.2,-0.3)$.
The ratio of disk height to radius is less than 0.25 at the arm positions.
These structures, such as the phase and height of the double-armed spirals, are consistent with the previously modeled disk height map of other WZ Sge-type DNe (V455 And; \cite{uem12ESHrecon}, OT J012059.6+325545; \cite{nak13j0120}).
Therefore, the early superhumps in PNV J0044 can be understood in the same manner as other WZ Sge-type DNe, even though its amplitude was the largest and the color trend was different from other WZ Sge-type DNe.

The unique point of PNV J0044 is that the inner arm structure is more evident than the other objects.
A closer look reveals that OT J012059.6+325545 does not have the lower-left inner arm structure compared to PNV J0044 and V455 And \citep{nak13j0120}.
In the case of V455 And, the height ratio to disk radius at the position of the inner arms is comparable to that of the outer disk \citep{uem12ESHrecon}.
However, the inner structure of PNV J0044 has a larger height ratio than its outer spiral arms. 
As these inner structures are hotter and bluer than the outer disk, this feature enables PNV J0044 not to be the reddest around the early superhump maximum, whereas other WZ Sge-type DNe show the reddest color around the early superhump maximum.
Even though this result can indicate that the inner disk in PNV J0044 is truly extended, another interpretation is possible;
as our model assumes the temperature $T$ gradient to the disk radius $R$ to be the standard disk model ($T \propto R^{-3/4}$; \cite{sha73alphadisk}), 
the inner elongated arm structure around $(X,Y)=(-0.3,-0.1)$ can mean a hotter temperature in the outer disk around  $(X,Y)=(-0.5,-0.1)$.
The same discussion can be applied to the upper right side of the disk.
Therefore, our results can also mean that the temperature of the outer disk in PNV J0044 and WZ Sge-type DNe is hotter than that of the standard disk model. 
An other independent method to test the temperature structure during the early superhump phase would be required to examine the disk temperature and height structure simultaneously.

\section{Summary}
\label{sec:5}

We report optical observations during the outburst of an eclipsing WZ Sge-type dwarf nova PNV J0044 in 2021.
Through the analysis of eclipses, its orbital period was determined as 0.055425534(1) d.
PNV J0044 showed early superhumps with the amplitude of 0.7 mag, which is the largest among known WZ Sge-type DNe.
This result proposes that PNV J0044 can be a WZ Sge-type DN with the highest inclination.
Moreover, its early superhumps showed the reddest color around the secondary minimum, whereas those of other well-observed WZ Sge-type DNe become the reddest around the primary maximum.
The spectra of PNV J0044 around the optical peak showed the double-peaked emission lines of H$\alpha$ and He II 4686\AA~ with the peak separation of $\ge$ 700 km s$^{-1}$, which supports the idea that PNV J0044 is a high-inclination system.
By applying the early superhump mapping to the multi-color and simultaneous observations with TriCCS, 
our result showed that the accretion disk of PNV J0044 in the early superhump phase accompanies a double-armed spiral structure. 
This result confirms that the large amplitude and unique color trend of the early superhumps in PNV J0044 originate from the 2:1 resonance as well as other WZ Sge-type DNe.

\begin{ack}

Y. T. acknowledges support from the Japan Society for the Promotion of Science (JSPS) KAKENHI Grant Number 21J22351. 
U.M., T.K., and D.N. acknowledge support from the JSPS KAKENHI Grant Number 21K03616.
This work was partially supported by the Slovak Research and Development Agency under the contract No. APVV-20-0148.
The authors thank the TriCCS developer team (which has been supported by the JSPS KAKENHI grant Nos. JP18H05223, JP20H00174, and JP20H04736, and by NAOJ Joint Development Research).

Lasair is supported by the UKRI Science and Technology Facilities Council and is a collaboration between the University of Edinburgh (grant ST/N002512/1) and Queen’s University Belfast (grant ST/N002520/1) within the LSST:UK Science Consortium. ZTF is supported by National Science Foundation grant AST-1440341 and a collaboration including Caltech, IPAC, the Weizmann Institute for Science, the Oskar Klein Center at Stockholm University, the University of Maryland, the University of Washington, Deutsches Elektronen-Synchrotron and Humboldt University, Los Alamos National Laboratories, the TANGO Consortium of Taiwan, the University of Wisconsin at Milwaukee, and Lawrence Berkeley National Laboratories. Operations are conducted by COO, IPAC, and UW. This research has made use of ``Aladin sky atlas'' developed at CDS, Strasbourg Observatory, France \cite{Aladin2000,Aladinlite}.
\end{ack}

\section*{Supporting Information}
The following Supporting Information is available on the online version of this article: Tables E1, E2 and E3.


\bibliographystyle{pasjtest1}
\bibliography{main}


\appendix

\end{document}